\documentclass[10pt,twocolumn]{revtex4-1}

\usepackage{amsmath, amsfonts, amssymb}     %
\usepackage{graphicx}                       %
\usepackage{xfrac}                          %
\usepackage{grffile}                        %
\usepackage{color,colortbl}

\usepackage[colorlinks, citecolor=blue, urlcolor=red, linkcolor=blue]{hyperref}

\newcommand{\figref}{Fig.~\ref}

\renewcommand{\eqref}{Eqn.~\ref}
\newcommand{\tsr}[1] {\mathbf{#1}}
\newcommand{\vtr}[1] {\mathbf{#1}}

\newcommand{\xhat}{\hat{\mathbf{x}}}
\newcommand{\yhat}{\hat{\mathbf{y}}}
\newcommand{\zhat}{\hat{\mathbf{z}}}

\newcommand{\gdot}[0] {\dot{\gamma}}

\newcommand{\degree}{\ensuremath{^\circ}}
\newcommand{\params}{\textbf{Parameters: }}

\newcommand{\vecv}[1]{\mathbf{{#1}}}
\newcommand{\tens}[1]{\mathbf{{#1}}}

\newcommand{\bi}{\begin{itemize}}
\newcommand{\ei}{\end{itemize}}

\newcommand{\be}{\begin{equation}}
\newcommand{\ee}{\end{equation}}
\newcommand{\bea}{\begin{eqnarray}}
\newcommand{\eea}{\end{eqnarray}}

\newcommand{\etas}{\eta_{\rm s}}
\newcommand{\etaa}{\eta_{\rm a}}

\begin{document}
\title{Edge fracture in complex fluids}
\author{Ewan J. Hemingway, Halim  Kusumaatmaja and Suzanne M. Fielding}
\affiliation{Department of Physics, Durham University, Science Laboratories,
  South Road, Durham DH1 3LE, UK}
\date{\today}
\begin{abstract}
  We study theoretically the edge fracture instability in sheared
  complex fluids, by means of linear stability analysis and direct
  nonlinear simulations. We derive an exact analytical expression for
  the onset of edge fracture in terms of the shear-rate derivative of
  the fluid's second normal stress difference, the shear-rate
  derivative of the shear stress, the jump in shear stress across the
  interface between the fluid and the outside medium (usually air),
  the surface tension of that interface, and the rheometer gap size.
  We provide a full mechanistic understanding of the edge fracture
  instability, carefully validated against our simulations.  These
  findings, which are robust with respect to choice of rheological
  constitutive model, also suggest a possible route to mitigating edge
  fracture, potentially allowing experimentalists to achieve and
  accurately measure stronger flows than hitherto.
\end{abstract}
\date{\today}
\maketitle

\begin{figure}[!b]
  \includegraphics[width=\columnwidth]{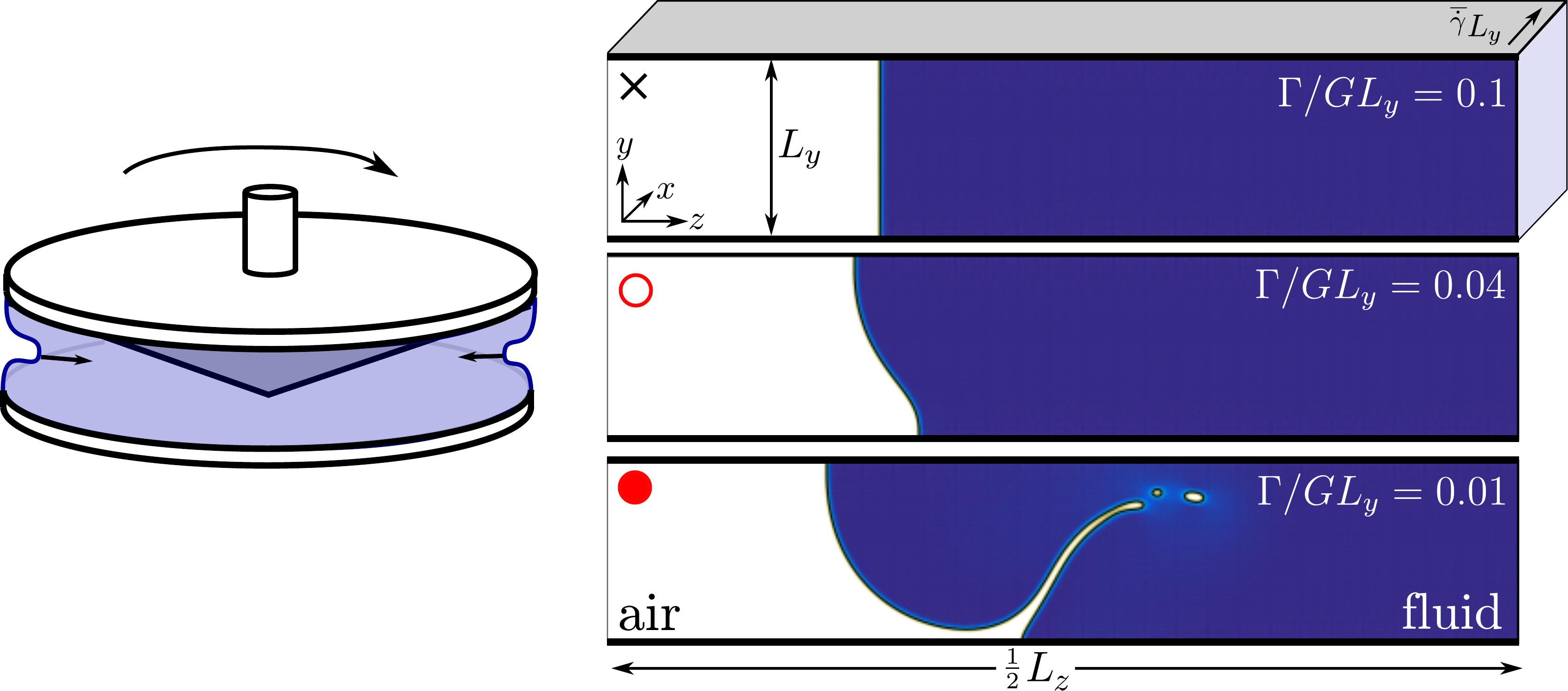}
  \caption{{\bf Left:} Schematic of a cone and plate device.
    {\bf Right:} Snapshots
  from full nonlinear simulations of the Giesekus model between hard
  walls.  $\gdot \tau=1.0$, $\theta=90^\circ$, $\alpha=0.4$,
  $\etaa / G \tau=0.01$. } \label{fig:geometry}
\end{figure}

Rheology is the study of the deformation and flow of matter. In the
most common rheological experiment, a sample of complex fluid -- eg.,
polymer, surfactant, colloid -- is sandwiched between plates and
sheared (Fig.~\ref{fig:geometry}).  Plotting the steady state shear
stress $\sigma$ as a function of imposed shear rate $\gdot$ then gives
the flow curve $\sigma(\gdot)$, which plays a central role in
characterising any fluid's flow response.  Almost ubiquitously
encountered beyond a certain (material and device dependent) shear
rate, however, is the phenomenon of edge fracture: the free surface
where the fluid sample meets the outside air destabilises
(Fig.~\ref{fig:geometry}, right), rendering accurate rheological
measurement impossible.  This has been studied experimentally in
Refs.~\cite{Lee1992,Inn2005,Sui2007,Schweizer2008,Mattes2008,Jensen2008,Dai2013}
and cited as ``the limiting factor in rotational
rheometry''~\cite{Jensen2008}. From a fluid mechanical viewpoint, it
is an important example of a hydrodynamic instability in free surface
viscoelastic flow~\cite{Eggers2008,Sui2007,Mckinley05viscoelasto}.

Despite this ubiquity, edge fracture remains poorly understood
theoretically. Important early papers by Tanner and
coworkers~\cite{Tanner1983,Keentok1999} predicted it to occur for a
critical magnitude $|N_2(\gdot)|>\Gamma/R$ of the second normal stress
difference $N_2$ in the fluid (we define $N_2$ below), given a surface
tension $\Gamma$ of the fluid-air interface and an assumed geometrical
lengthscale $R$. This prediction was based on some key assumptions
that will in fact prove inconsistent with our simulations.  Taken as a
scaling argument, however, it showed remarkable early insight.

The contributions of this Letter are fourfold. First, we show that the
threshold for the onset of edge fracture is in fact set by $\Delta
\sigma \; |N_2|'(\gdot)\,/\,\sigma'(\gdot)>2\pi\Gamma/L_y$, where
prime denotes differentiation with respect to $\gdot$, $\Delta\sigma$
is the jump in shear stress across the interface between the fluid and
the outside air, and $L_y$ is the gap size. (For a note on signs,
see~\cite{Note1}.)  For low flow rates and negligible air viscosity,
setting also $R=L_y$, Tanner's prediction happens to equal ours to
within an $O(1)$ factor, despite containing fundamentally different
physics. Second, we offer the first mechanistic understanding of edge
fracture. Third, we predict the growth rate at which it develops for
any imposed shear rate.  Finally, we suggest a recipe by which it
might be mitigated, potentially enabling experimentalists to achieve
stronger flows than hitherto.

Our approaches comprise linear stability analysis and direct nonlinear
simulation. At low shear rates in a simplified theoretical
geometry~\cite{Onuki1997a}, defined below, we obtain exact
expressions for the threshold, eigenvalue and eigenfunction for the
onset of edge fracture, and show these to agree with counterpart
nonlinear simulations. We further show this simplified geometry to
closely predict onset in the experimentally realisable geometry of
shear between plates.

As shown in Fig.~\ref{fig:geometry} (right), we consider a planar slab
of fluid sheared at rate $\gdot$ with flow direction $\xhat$ and
flow-gradient direction $\yhat$. For a small cone angle and large
radius in the flow cell sketched in Fig.~\ref{fig:geometry}, left,
which is usually the case experimentally, this planar cartoon provides
an excellent approximation. The edges of the sample in the vorticity
direction $\zhat$ are in contact with the air, with a sample length in
that direction (initially, at the cell midheight $y=0$) denoted
$\Lambda$. We assume translational invariance in $\xhat$, performing
two-dimensional simulations in the $y-z$ plane. Our simulation box has
length $L_z$ and periodic boundary conditions in $z$. Only its left
half is shown in Fig.~\ref{fig:geometry}.

In the $y$ direction we consider two different kinds of boundary
condition.  The first models the experimentally realisable case of
shear between hard walls at $y=\pm L_y/2$, with no slip or
permeation. The second gives the simplified biperiodic Lees-Edwards
geometry, in which all quantities repeat periodically across
shear-mapped points on the boundaries of box copies stacked in $y$,
but with adjacent copies moving relative to each other at velocity
$\gdot L_y \hat{\vtr{x}}$. Our numerically obtained threshold for the
onset of edge fracture will prove in excellent agreement between these
two. The simplified geometry allows analytical progress that is
otherwise prohibitive.

The total stress $\tens{T}$ in any fluid element comprises an
isotropic contribution $-p\tens{I}$ with pressure $p$, a Newtonian
solvent contribution of viscosity $\etas$, and a viscoelastic
contribution $\tens{\Sigma}$ from the complex fluid (polymer chains,
emulsion droplets, etc.), with a scale set by a constant modulus $G$.
We assume creeping flow conditions, giving the force balance condition
$\nabla.\tens{T}=0$, and therefore
$\etas\nabla^2\vecv{v}+\nabla.\tens{\Sigma}-\nabla p=0$ inside the
fluid and $\etaa\nabla^2\vecv{v}-\nabla p=0$ in the air, with air
viscosity $\etaa$. The pressure field $p(\vecv{r},t)$ is determined by
enforcing incompressibility, with the flow velocity
$\vecv{v}(\vecv{r},t)$ obeying $\nabla.\vecv{v}=0$.  The dynamics of
$\tens{\Sigma}$ is determined by a viscoelastic constitutive equation
of the form
\be
\partial_t\tens{\Sigma}+\vecv{v}.\nabla\tens{\Sigma} = 2G\tens{D} +
\tens{f}(\tens{\Sigma},\nabla\vecv{v})-\frac{1}{\tau}\tens{g}(\tens{\Sigma}),
\label{eqn:vece}
\ee
where $\tens{D}=\tfrac{1}{2}(\nabla \vecv{v} + \nabla \vecv{v}^T)$.
The first two terms on the RHS capture the loading of viscoelastic
stress in flow; the third relaxation back towards an unstressed
state. The forms of $\tens{f}$ and $\tens{g}$ prescribe the precise
model, and we shall simulate in what follows the
Johnson-Segalman~\cite{Johnson1977} and Giesekus~\cite{Giesekus1982}
models, set out in~\cite{SI}. In the former, $\tens{f}$ contains a
slip parameter $a$. In the latter, $\tens{g}$ contains an anisotropy
parameter $\alpha$.  Importantly, however, our predictions for edge
fracture will depend on $a$ or $\alpha$ only via their appearance in
the shear stress $\sigma\equiv T_{xy}$ and second normal stress
difference $N_2\equiv T_{yy}-T_{zz}$. In this way, the key physics
proves robust to choice of constitutive model. Indeed, most complex
fluids show the low-shear scalings $\sigma\sim\gdot,N_2\sim-\gdot^2$
of this model. An exception are non-Brownian suspensions~\cite{Denn2014}, deferred to
future work.

Our simulations model the air-fluid coexistence by a Cahn-Hilliard
equation~\cite{SI,Anderson1998,Kusumaatmaja2016}, with a mobility $M$ for
air-fluid intermolecular diffusion, a scale $G_\mu$ for the free
energy density of demixing, and a slightly diffuse air-fluid interface
of thickness $l$, with surface tension $\Gamma=2\sqrt{2}G_\mu l/3$.
Our linear stability analysis assumes a sharp interface, with a
surface tension $\Gamma$. Our results for these two approaches agree
fully.

\begin{figure}[t]
  \includegraphics[width=\columnwidth]{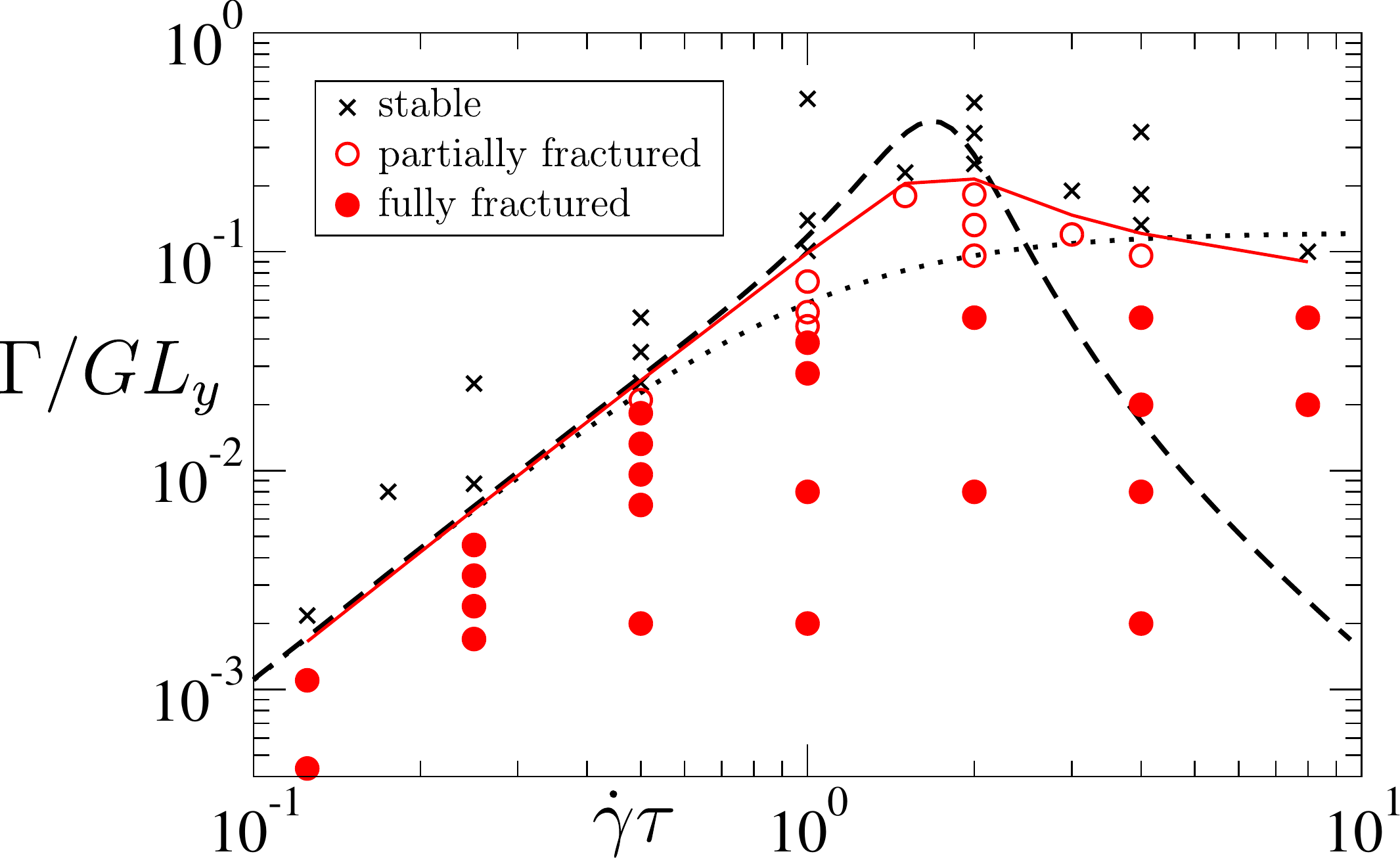} \caption{Edge fracture
  phase diagram for the Johnson-Segalman model in Lees-Edwards
  biperiodic shear.  Solid line: phase boundary between stable and
  partially fractured states. Dashed line: prediction
  of~\eqref{eqn:criterion}, with no adjustable parameters.
  Dotted line: Tanner's prediction, with the prefactor adjusted to best-fit
the simulations in the limit $\gdot \tau \to 0$. $a = 0.3$, $\etaa / G\tau = 0.01$. }
  \label{fig:PD}
\end{figure}

In unsheared equilibrium, the contact angle where the air-fluid
interface meets the flow cell walls is denoted $\theta$. A value
$\theta=90^\circ$ gives a vertical equilibrium interface;
$\theta>90^\circ$ an interface convex into the air; and
$\theta<90^\circ$ concave. In having a diffuse interface~\cite{Kusumaatmaja2016}, our
simulations capture any motion of the contact line along the
wall in flow. In the simplified biperiodic
geometry the equilibrium interface is always vertical, mimicking
$\theta=90^\circ$ with walls. As the initial condition for our shear
simulations, we take a coexistence state first equilibrated without
shear, with a small perturbation then added to the interface's
position $h(y)$ along the $z$ axis, $h\to h+10^{-8}\cos(n\pi y/L_y)$,
to trigger edge fracture, taking $n=1$ with walls and $n=2$ in the
biperiodic geometry.

Important dimensionless quantities that we
shall explore are the scaled surface tension
$\Gamma/GL_y$, the Weissenberg number $\gdot\tau$, the
equilibrium contact angle $\theta$, and the air viscosity
$\etaa/G\tau$. Less important parameters, which do not affect the
physics once converged to their physically appropriate large or small
limit are: the cell aspect ratio, $L_z/L_y=10.0$; the air gap size
$(L_z-\Lambda)/L_y=3.0$; the small solvent viscosity~\cite{SI};
the air-fluid interface width $l/L_y=0.01$, and the inverse mobility
for intermolecular diffusion, $l^2/MG_\mu \tau=0.01-0.1$.

We now present our results. The basic phenomenon is exemplified by the
three late-time snapshots of our nonlinear simulations of the Giesekus
model between hard walls in Fig.~\ref{fig:geometry}, right.  At any
given imposed strain rate, an air-fluid interface with high surface
tension is undisturbed by the flow and retains its equilibrium shape
(top snapshot). We shall denote such states by a black cross in
Fig.~\ref{fig:PD}. For an intermediate surface tension the interface
partially fractures, displacing in the $z$ direction a distance
$O(L_y)$ set by the gap between the rheometer plates in the $y$
direction, before settling to a new steady state shape, different from
its unsheared equilibrium one.  We denote these states by red open
circles. Finally for a low surface tension, the interface fully
fractures, displacing in the $z$ direction a distance $O(\Lambda)$ set
by the sample width in that direction (red closed circles). Here the
system never attains a new steady state: depending on the wetting
angle and flow rate, the fluid may, eg, de-wet the wall, and/or air
bubbles invade the fluid.

In Fig.~\ref{fig:PD}, we collect into a phase diagram the results of
simulations at many values of surface tension and shear rate, for the
Johnson-Segalman model in the biperiodic geometry. (In the
SI~\cite{SI}, we show that the phase boundary is essentially
independent of model, geometry and equilibrium wetting angle
$\theta$.) The red solid line marks the phase boundary between
undisturbed and partially fractured interfacial states.

\begin{figure}[t]
  \includegraphics[width=\columnwidth]{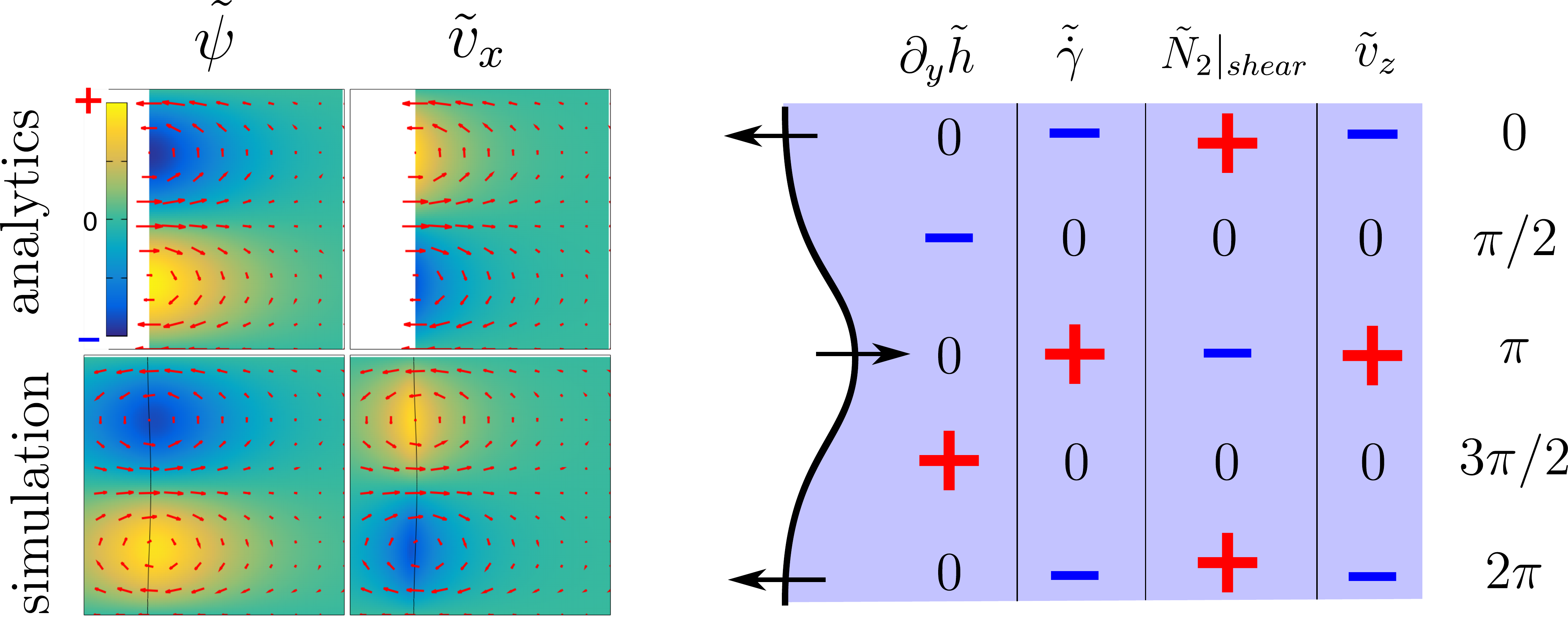} \caption{{\bf
  Left:} eigenfunctions from analytic calculation (top) and simulation
  (bottom). (Analytics ignore the air phase, as shown by the white regions.) $\gdot\tau =
  0.125$, $a=0.3$, $\Gamma / G L_y = 0.0$, $\etaa/G\tau = 0.01$, $q=2\pi/L_y$. {\bf
Right:} instability mechanism, discussed in text, with $0$ and $\pm$ symbols corresponding to the phase locations shown.}
  \label{fig:eigenfunction}
\end{figure}

Within the biperiodic geometry, we now perform a linear stability
analysis to derive an expression for this onset threshold, in the
limit of low strain rates.  To do so, we represent the state of the
system as an underlying homogeneous time-independent base state
(denoted by subscript $0$), corresponding to the initially unfractured
case in which the interface is flat and the flow uniform. (Recall that
our nonlinear simulations showed the phase boundary to be independent
of the initial interfacial shape~\cite{SI}.)  To this, we add a small
perturbation (denoted by over-tildes) representing the precursor of
edge fracture.  For any given interfacial tension $\Gamma$ and imposed
flow rate $\gdot$, we then determine whether the perturbation grows
towards an edge fractured state, or decays to leave a flat interface.

Accordingly, in the fluid bulk we write the velocity field
$\vecv{v}=\vecv{v}_0+\tilde{\vecv{v}}=(\gdot
y,0,0)+(\tilde{v}_x,\partial_z\tilde{\psi},-\partial_y\tilde{\psi})$
and stress field $\tens{T}=\tens{T}_0+\tilde{\tens{T}}$. Our use of a
streamfunction $\tilde{\psi}$ automatically ensures
incompressibility. The force balance condition $\nabla.\tens{T}=0$
then simply becomes $\nabla.\tilde{\tens{T}}=0$.  In the fluid bulk
the $x$ component of force balance, and the curl of its $y,z$
components are respectively:
\begin{subequations}
\bea
0&=&\partial_y \tilde{T}_{xy}+\partial_z\tilde{T}_{xz},\label{eqn:bulk_a}\\
0&=&\partial_y\partial_z(\tilde{T}_{yy}-\tilde{T}_{zz})+(\partial_z^2-\partial_y^2)\tilde{T}_{yz}.\label{eqn:bulk_b}
\eea
\label{eqn:bulk}
\end{subequations}
We likewise write the $z$-position of the interface at any gap
coordinate $y$ as $h_0+\tilde{h}(y)$. We further choose the origin of
$z$ to lie at the interface, so $h_0=0$, with fluid for $z>0$ and air
for $z<0$. The condition of force balance
$\vecv{n}.\tens{T}+\Gamma\vecv{n}\nabla_{\rm int}.\vecv{n}=0$ across
this perturbed interface with normal
$\vecv{n}=\zhat-\partial_y\tilde{h}\yhat$ and $\nabla_{\rm int}$
the interfacial gradient operator gives componentwise
linearised equations:
\begin{subequations}
\bea
0&=&\tilde{T}_{xz}|_{z=0^+}-\Delta \sigma\,\,\partial_y h,\label{eqn:interface_a}\\
0&=&\tilde{T}_{yz}|_{z=0^+}- N_2 \,\partial_y h,\label{eqn:interface_b}\\
0&=&\tilde{T}_{zz}|_{z=0^+}+\Gamma\,\partial_y^2h,\label{eqn:interface_c}
\eea
\label{eqn:interface}
\end{subequations}
\noindent with $\Delta\sigma$ and $N_2$ the jumps in the shear and second normal
stress difference across the interface, from fluid to air. ($N_2$ is
always zero in the air, so we omit its $\Delta$ prefix.)  Note we have
assumed (for now) negligible stresses on the air side of the
interface, $z=0^-$.  The interface moves with the $z$-component of the
fluid velocity:
\be
\partial_t\tilde{h}=-\partial_y\tilde{\psi}|_{z=0}.
\label{eqn:moves}
\ee

Finally, we must specify the perturbed stress components
$\tilde{T}_{ij}$ in Eqns.~\ref{eqn:bulk} and~\ref{eqn:interface}. Each
comprises a solvent contribution of viscosity $\etas$, and a
viscoelastic stress that follows Eqn.~\ref{eqn:vece}. For values of
$(\Gamma / G L_y,\gdot\tau)$ only just across the instability threshold in
Fig.~\ref{fig:PD}, the interface will destabilise only very slowly and
the viscoelastic stress will, for any instantaneous interfacial shape,
be determined as the quasistatic solution of Eqn.~\ref{eqn:vece}.  In
the limit of small imposed shear rate $\gdot$, this gives
\begin{subequations}
\bea
\tilde{T}_{xy}&=&(G\tau+\etas)\partial_y\tilde{v}_x+O(\gdot),\\
\tilde{T}_{xz}&=&(G\tau+\etas)\partial_z\tilde{v}_x+O(\gdot),\\
\tilde{T}_{yy}-\tilde{T}_{zz}&=&4(G\tau+\etas)\partial_y\partial_z\tilde{\psi} -2\gdot G\tau^2 b\partial_y\tilde{v}_x,\label{eqn:components_c}
\eea
\label{eqn:components}%
\end{subequations}
\noindent with $b=1-a$ and $\alpha$ in the Johnson-Segalman and Giesekus models
respectively.

Substituting Eqn.~\ref{eqn:components} (with a counterpart expression for $\tilde{T}_{yz}$) into
Eqns.~\ref{eqn:bulk}, \ref{eqn:interface} gives
finally a set of coupled partial differential equations for the
perturbation to the bulk flow field,
$\tilde{v}_x(y,z,t),\tilde{\psi}(y,z,t)$, and to the interface
position $\tilde{h}(y,t)$. Solving these gives, to leading order in
$\gdot$ and at any wavevector $q$ in the $y$ direction,
\bea
\tilde{\psi}(y,z,t)&=&\left[A e^{-qz}+B e^{-kz}\right]e^{iqy}e^{\omega t},\nonumber\\
\tilde{v}_x(y,z,t)&=&C e^{-qz}e^{iqy}e^{\omega t},\nonumber\\
\tilde{h}(y,t)&=&iq D e^{iqy}e^{\omega t},\label{eqn:eigenfunction}
\eea
(ignoring a small term in $e^{-kz}$ in $\tilde{v}_x$), in which
$k=q/\sqrt{1+\beta}$ with $\beta \approx b(1-b)\gdot^2 \tau^2$, and with
known expressions for $A,B,C,D$ that we do not write.  These
eigenfunctions $\tilde{\psi}(y, z)$, $\tilde{v}_x(y, z)$ are
shown in the left panel of Fig.~\ref{fig:eigenfunction}
and agree fully with their counterparts from (the linear regime of)
our fully nonlinear simulations in the same panel.

\begin{figure}[t]
  \includegraphics[width=\columnwidth]{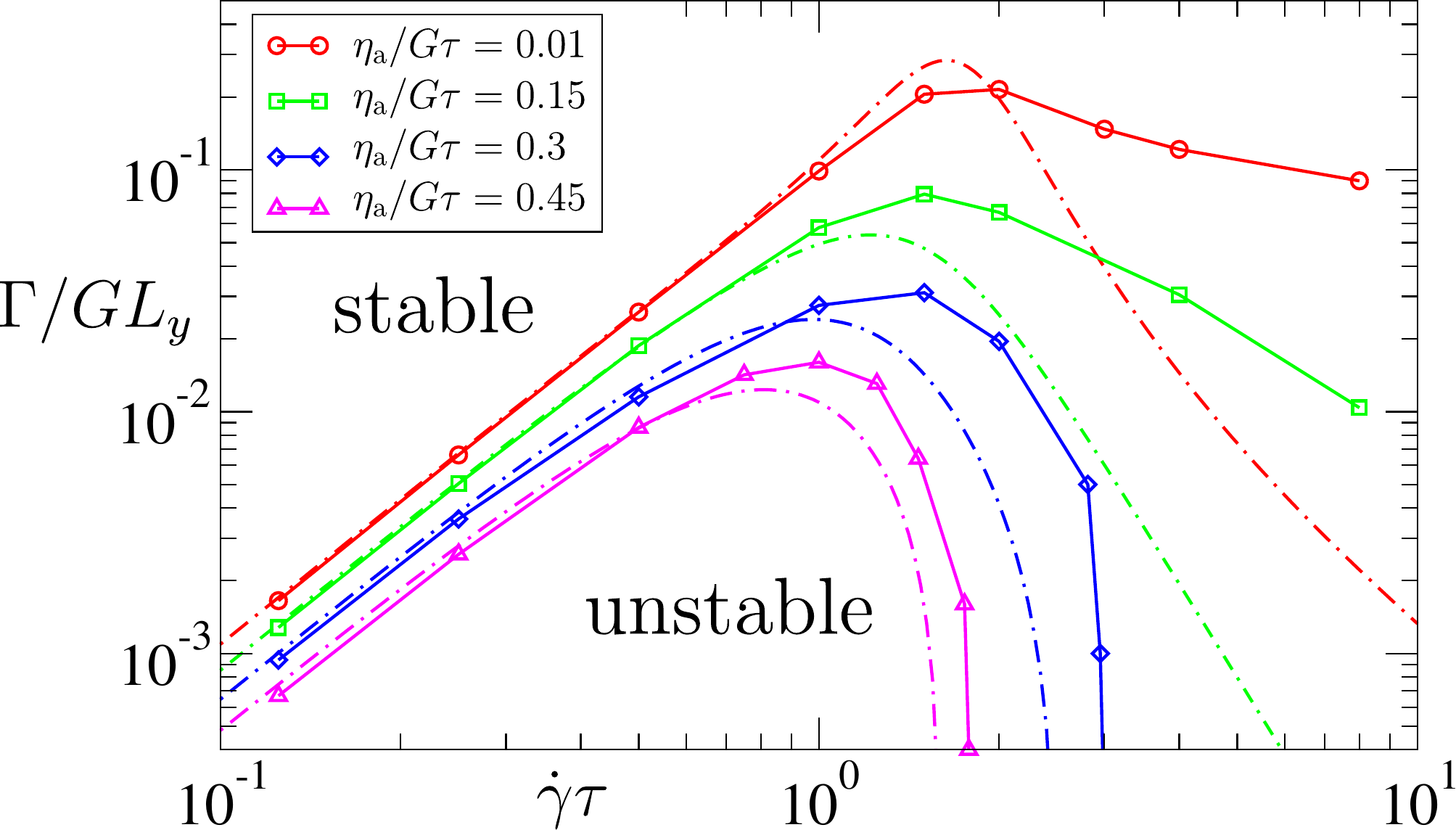}
  \caption{Threshold for onset of edge fracture instability in the
  Johnson-Segalman model in biperiodic shear, for various values of
  the viscosity $\etaa$ of the bathing medium. Solid lines: full
  nonlinear simulation. Dotted lines: linear stability analysis, valid
  in the limit $\gdot\tau \to 0$.  $a = 0.3$, $\etas/G\tau = 0.15$.  }
  \label{fig:eta_air_PD}
\end{figure}

Eqn.~\ref{eqn:eigenfunction} tells us that perturbations at any
wavevector $q$ will grow if their eigenvalue $\omega(q)>0$. We find
\be
\omega=\frac{1}{2(G\tau+\etas)}\left[\frac{1}{2}\Delta\sigma\frac{d|N_2|}{d\gdot}\middle/\frac{d\sigma}{d\gdot}-\Gamma q\right].
\label{eqn:eigenvalue}
\ee
The condition $\omega>0$ is most readily satisfied for the mode with
the lowest wavevector that is consistent with the boundary conditions,
$q=2\pi/L_y$.  Accordingly, our final condition for an initially flat
fluid-air interface to undergo edge fracture is given by
\be
\frac{1}{2}\Delta\sigma\frac{d|N_2(\gdot)|}{d\gdot}\bigg/\frac{d\sigma}{d\gdot}>\frac{2\pi\Gamma}{L_y}.
\label{eqn:criterion}
\ee
This criterion is marked by the dashed line in Fig.~\ref{fig:PD}, and
fully agrees at low shear rates with the onset of fracture in our
numerical simulations.

We now compare Eqn.~\ref{eqn:criterion} with Tanner's prediction of
$|N_2|>2\Gamma/3R$, with $R$ the radius of an assumed initially
semicircular interfacial crack. Clearly, $R$ must now be replaced by
the dominant wavelength $L_y$.  Disregarding $O(1)$ prefactors, the
important difference between Tanner's prediction and ours then lies in
replacing
\be
|N_2|\to\frac{1}{2}\;\Delta\sigma\frac{d|N_2|}{d\gdot}\bigg/\frac{d\sigma}{d\gdot}.\;\;
\label{eqn:replace}
\ee
Given negligible air viscosity, the jump $\Delta\sigma$ in shear
stress across the interface between the fluid and air simply equals
the shear stress $\sigma$ in the fluid.  For most complex fluids
(excluding non-Brownian suspensions), in the limit of small shear
rates, $N_2\sim -\gdot^2$ and $\sigma\sim\gdot$. Tanner's $|N_2|$ on
the LHS of (\ref{eqn:replace}) then simply equals our expression on
the RHS.  In contrast, at higher shear rates these simple power laws
no longer (in general) hold, and our prediction departs from Tanner's,
as seen in Fig.~\ref{fig:PD}. Indeed, Tanner predicts the critical
surface tension to increase monotonically with shear rate. The
non-monotonicity that we find follows because $\sigma$ and $|N_2|$
both initially increase with $\gdot$, before $N_2(\gdot)$ saturates to
a constant at high shear rates, such that $dN_2/d\gdot\to 0$.

Our results also explain the mechanism of instability as follows.  Were the
interface to remain perfectly flat, the jump $\Delta\sigma$ in shear
stress across it would be consistent with force balance.  However, any
small interfacial tilt $\partial_y\tilde{h}$ (first column of
Fig.~\ref{fig:eigenfunction}, right) exposes this jump. To maintain
force balance across the interface, a counterbalancing perturbation
$\tilde{T}_{xz}=iq h\Delta\sigma$ is then required
(Eqn.~\ref{eqn:interface_a}).  To maintain the $x$-component of force
balance in the fluid bulk (Eqn.~\ref{eqn:bulk_a}), a corresponding
perturbation $\tilde{T}_{xy}$ is then needed, achieved via a
perturbation $\tilde{\gdot}=\partial_y\tilde{v}_x=qh\Delta
\sigma/\sigma'(\gdot)$ in the shear rate (second column of
Fig.~\ref{fig:eigenfunction}, right).  The second normal stress $N_2
\approx -bG\tau^2\gdot^2$
in the fluid bulk then suffers a corresponding perturbation (second
term in Eqn.~\ref{eqn:components_c})
$\tilde{T}_{yy}-\tilde{T}_{zz}|_{\rm shear}=-qh\Delta\sigma\,
|N_2|'(\gdot)/\sigma'(\gdot)$ (third column of
Fig.~\ref{fig:eigenfunction}, right). This must be counterbalanced (at
zero surface tension at least) by an equal and opposite extensional
perturbation (first term in Eqn.~\ref{eqn:components_c}):
$\tilde{T}_{yy}-\tilde{T}_{zz}|_{\rm
ext}=4G\tau\partial_y\partial_z\tilde{\psi}=-4G\tau\partial_z\tilde{v}_z=4G\tau q\tilde{v}_z$. This
requires a $z$-component of fluid velocity (fourth column of
Fig.~\ref{fig:eigenfunction}, right), which convects the interface,
$\partial\tilde{h}/\partial t=\tilde{v}_z=\tfrac{1}{4}\Delta\sigma
h\,|N_2|'(\gdot)/G\tau \sigma'(\gdot)$, enhancing its original tilt with a
growth rate
$\omega=\tfrac{1}{4}\Delta\sigma\,|N_2|'(\gdot)/G\tau \sigma'(\gdot)$,
consistent with Eqn.~\ref{eqn:eigenvalue} at zero surface tension,
noting that $\etas$ is small. This mechanism resembles in spirit that of
instabilities between layered viscoelastic
fluids~\cite{Hinch1992,Wilson1997,Fielding2010}.

Finally, our results suggest a recipe via which edge fracture might be
mitigated. By immersing the flow cell in an immiscible Newtonian
`bathing fluid' with a viscosity larger than that of air, more closely
matched to that of the study-fluid, the jump $\Delta\sigma$ in shear
stress between the study and bathing fluids, which is a key factor in
driving the instability, will be reduced. This is explored in
Fig.~\ref{fig:eta_air_PD}. The red solid line shows the onset
threshold for a bathing fluid of negligible viscosity, such as air;
and the green, blue and magenta lines for successively increasing
values of the bathing fluid's viscosity, each giving increased
stability. The dashed lines show linear stability results recalculated
with non-zero bath viscosity, in excellent agreement.  Clearly, choosing
a bathing fluid with as a high a possible surface tension with the test
fluid will also help stability.

To summarise, we have derived an exact expression for the onset of
edge fracture in complex fluids, shown it to agree with numerical
simulations, and provided the first mechanistic understanding of edge
fracture. We have also suggested a way of mitigating the phenomenon
experimentally. Given the status of edge fracture as a crucially
limiting factor in experimental rheology, this suggests a route to
accessing stronger flows than hitherto.

{\it Acknowledgements -- } The research leading to these results has
received funding from the European Research Council under the EU's 7th
Framework Programme (FP7/2007-2013) / ERC grant number 279365. The
authors thank Peter Olmsted for discussions and Mike Cates and Roger Tanner
for a critical reading of the manuscript.

%

\pagebreak
\clearpage
\begin{center}
\textbf{\large Supplementary Material for:\\ ``Edge fracture in complex fluids''}
\end{center}
\setcounter{equation}{0}
\setcounter{figure}{0}
\setcounter{table}{0}
\makeatletter
\renewcommand{\theequation}{S\arabic{equation}}
\renewcommand{\thefigure}{S\arabic{figure}}
\renewcommand{\bibnumfmt}[1]{[S#1]}
\renewcommand{\citenumfont}[1]{S#1}


This supplementary information is divided into four parts. In the
first, we define the details of the constitutive models for which
results are presented in the main text. In the second, we show those
results to be independent of this choice of constitutive model. In the
third, we show robustness to the boundary conditions at edges of the
flow cell. Finally, we outline our numerical scheme.

\section{Definition of constitutive models}

As discussed in the main text, the dynamics in flow of the
viscoelastic stress $\tens{\Sigma}$ is determined by a
constitutive equation of the general form
\be
\partial_t\tens{\Sigma}+\vecv{v}.\nabla\tens{\Sigma} = 2G\tens{D} +
\tens{f}(\tens{\Sigma},\nabla\vecv{v})-\frac{1}{\tau}\tens{g}(\tens{\Sigma}),
\label{eqn:SI_vece}
\ee
where $G$ is the viscoelastic modulus and $\tau$ is the relaxation timescale.
The forms of $\tens{f}$ and $\tens{g}$ depend on the
constitutive model in question, and we now specify these for the two
models explored in the main text.

The Johnson-Segalman model~\cite{Johnson1977a} has
\begin{align}
  \tsr{f}\left(\tsr{\Sigma}, \tsr{\nabla v}\right) &= \left(\tsr{\tsr{\Sigma} \tsr{\Omega} - \Omega} \tsr{\Sigma} \right) + a\left(\tsr{D} \tsr{\Sigma} + \tsr{\Sigma} \tsr{D}\right),\\
  \tsr{g}\left(\tsr{\Sigma}\right) &= \tsr{\Sigma},
  \label{eq:SI_JS_f_g}
\end{align}
in which $\tens{D}=\tfrac{1}{2}(\nabla \vecv{v} + \nabla \vecv{v}^T)$ and
$\tens{\Omega}=\tfrac{1}{2}(\nabla \vecv{v} - \nabla \vecv{v}^T)$ with
$\nabla \vecv{v}_{\alpha\beta}=\partial_\alpha v_\beta$.  The
parameter $a$ describes the slip of the viscoelastic component (eg,
polymer chains) relative to affine flow. It has values in the range
$-1 \le a\le 1$. Define the adimensional viscoelastic shear stress,
second normal stress difference and shear rate as
\begin{equation}
  \hat{\sigma}_{\rm p} = \Sigma_{xy} / G, \quad \hat{N}_2 = \left(\Sigma_{yy} - \Sigma_{zz}\right)/G,\quad\hat{\gdot} = \gdot \tau
  \label{eq:adimensionalised}
\end{equation}
respectively. Then in steady homogeneous simple shear flow, as
function of the imposed shear rate $\gdot$, these are given
as~\cite{Skorski2011}
\begin{align}
  \hat{\sigma}_{\rm p} = \frac{\hat{\gdot}}{1 + (1 - a^2) \hat{\gdot}^2},\\
  \hat{N}_2 = \frac{\left(-1 + a\right) \hat{\gdot}^2}{1 + (1 - a^2) \hat{\gdot}^2}.
  \label{eq:sigma_N2_JS}
\end{align}

The Giesekus model~\cite{Giesekus1982a} has
\begin{align}
  \tsr{f}\left(\tsr{\Sigma}, \tsr{\nabla v}\right) &= \left(\tsr{\tsr{\Sigma} \tsr{\Omega} - \Omega} \tsr{\Sigma} \right) + \left(\tsr{D} \tsr{\Sigma} + \tsr{\Sigma} \tsr{D}\right),\\
  \tsr{g}\left(\tsr{\Sigma}\right) &= \tsr{\Sigma} + \frac{\alpha}{G} \tsr{\Sigma}^2.
  \label{eq:SI_Gies_f_g}
\end{align}
Here $\alpha$ is an anisotropy parameter, which models an enhanced
rate of stress relaxation in regimes where the polymer chains are more
strongly aligned. In steady homogeneous simple shear flow, the adimensional
viscoelastic shear stress and second normal stress difference are
given as~\cite{Giesekus1982a}
\bea
\hat{\sigma}_{\rm p}&=&\frac{(1+\hat{N}_2)^2\hat{\gdot}}{1-(1-2\alpha)\hat{N}_2},\nonumber\\
\hat{N}_2&=&\frac{\left(\Lambda - 1\right)}{1+(1-2\alpha)\Lambda},
\eea
in which
\be
\Lambda^2=\frac{1}{8\alpha(1-\alpha)\hat{\gdot}^2}\left[ \sqrt{1+16\alpha(1-\alpha)\hat{\gdot}^2}-1  \right].
\ee

In both models, the total steady state shear stress
$\sigma(\gdot)=G\hat{\sigma}_{\rm p}(\gdot)+\etas\gdot$ comprises the sum of
the viscoelastic part just defined and a Newtonian contribution of
viscosity $\etas$. For parameter values $|a|<1$ and $\etas / G\tau < 1/8$ in
the Johnson-Segalman model, the total shear stress $\sigma(\gdot)$ is
a non-monotonic function of the imposed shear rate, allowing the
coexistence of bands of differing shear rates at a common value of the
total shear stress: a phenomenon known as shear banding.  We consider
here only non-shear-banded flows, taking $a=0.3$ and $\etas / G\tau=0.15$ in
our numerical simulations. In the Giesekus model, $\sigma(\gdot)$ is a
monotonic function for $\alpha\le1/2$ for any $\etas$
\cite{Giesekus1982a}. We set $\alpha=0.4$ and $\etas/G\tau = 0.01$ in our numerics,
again avoiding shear-banding. In both models (provided $a<1$ or
$\alpha>0$), the second normal stress is negative, scaling as
$-\gdot^2$ at low shear rates and saturating to a negative constant at
high shear rates.

\section{Robustness to constitutive model}

\begin{figure}[!t]
  \includegraphics[width=\columnwidth]{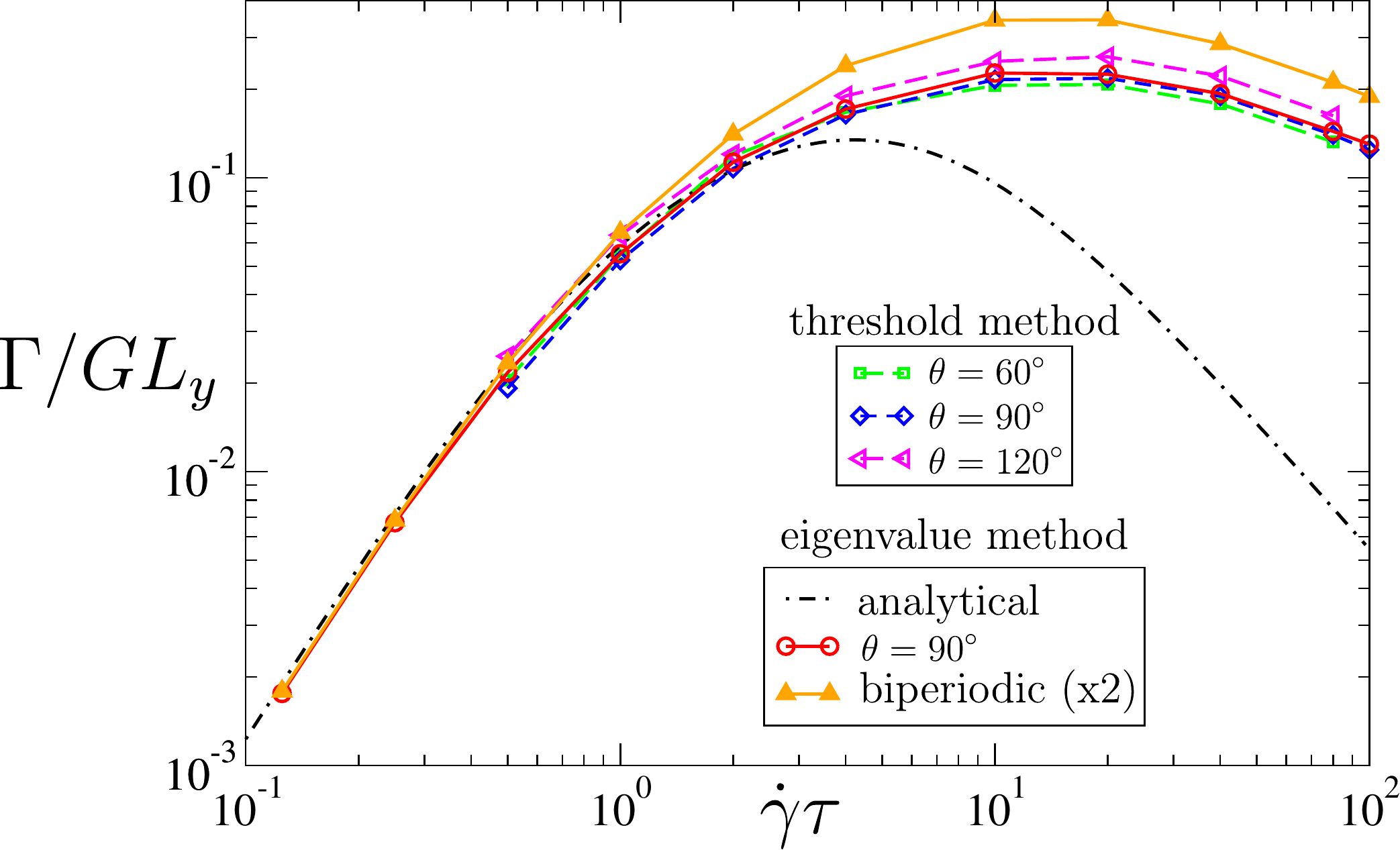}
  \caption{Threshold for the onset of edge fracture instability in  the
  Giesekus model sheared between hard walls (open symbols) and in the
  biperiodic geometry (closed symbols). Black dash-dotted line shows our
  analytical prediction of Eqn. 8 in the main text.  \params
  $\alpha=0.4$, $\etas / G\tau =0.01$, $\etaa / G\tau = 0.01$. }
  \label{fig:BCs}
\end{figure}

We now demonstrate the robustness of our results with respect to this
choice of constitutive model. Recall that Fig.~2 of the main text
showed the threshold for the onset of edge fracture obtained from our
numerical simulations of the Johnson-Segalman model in the biperiodic
geometry, and the agreement with it at low strain rates of our
analytical prediction of Eqn. 8 of the main text. We now explore this
same comparison for the Giesekus model. See \figref{fig:BCs}, in which
the solid triangles show the threshold obtained from our numerical
simulations in the biperiodic geometry, and the long-dashed line shows
the prediction of Eqn. 8 of the main text. Excellent agreement is
again obtained at low shear rates. Both constitutive models therefore
give behaviour in accordance with our central prediction of Eqn. 8 in
the main text.

Comparing~\figref{fig:BCs} with Fig. 2 of the main text also shows
that both models capture a re-entrant region of stability against edge
fracture at large shear-rates. As discussed in the main text, this
arises from the saturation of $N_2$ at high shear rates. The smaller
solvent viscosity in the Giesekus simulations however postpones this
to higher strain rates than in the Johnson-Segalman simulations.

\section{Robustness to boundary condition}

Recall that in the main text we considered two different boundary
conditions: the first corresponding to the experimentally realisable
geometry in which the boundaries of the flow cell in the flow-gradient
direction comprise hard walls, and the second to theoretically
simplified Lees-Edwards sheared periodic boundary conditions.

To check for robustness with respect to this choice of boundary
condition, in~\figref{fig:BCs} we compare the threshold for the onset
of edge fracture obtained from simulations of a cell with hard walls
in the flow-gradient direction, for three different values of the
equilibrium contact angle, $\theta = 60, 90, 120\degree$, with that
obtained in simulations adopting Lees-Edwards biperiodic shear. Good
qualitative agreement is seen across these four cases at all shear
rates, with excellent quantitative agreement at low shear rates. (Note
that the lowest possible wavevector in the biperiodic geometry is
twice that in the walled geometry. For consistency we accordingly
rescaled the critical surface tension by a factor two in that case.)

To identify this threshold, we first defined the steady-state
displacement of the interface to be $d =
\max(h(y)) - \min(h(y))$, with $d_0$ the value of this quantity in an
unsheared system. We then define the onset threshold at any imposed
shear rate to be the value of the surface tension $\Gamma$ at which
$d(\Gamma)$ (in shear) obeys $d(\Gamma) - d_0 = 0.1$. For a contact
angle $\theta=90\degree$ in the simulations with walls, and in all the
simulations in biperiodic shear, the interface between the fluid and
air is initially flat and we can alternatively identify onset of edge
fracture by the surface tension at which the eigenvalue calculated in
the main text first becomes positive. As seen in Fig.~\ref{fig:BCs},
these two methods of identifying onset agree well.

\section{Numerical scheme}

In our analytical calculations we assume an infinitely sharp interface
of surface tension $\Gamma$ between the sheared slab of viscoelastic
fluid and the outside air. In our simulations we instead explicitly
model this coexistence of fluid and air using a phase field approach
with an order parameter $\phi$, which obeys Cahn-Hilliard
dynamics~\cite{Bray1994}
\begin{align}
  \partial_t \phi + \vecv{v}.\nabla \phi &= M \nabla^2 \mu.
  \label{eq:SI_CH}
\end{align}
Here $M$ is the molecular mobility, which we assume constant. The
chemical potential
\begin{align}
  \mu &= G_\mu \left(-\phi + \phi^3 - \ell^2 \nabla^2 \phi\right),
\end{align}
in which $G_\mu$ sets the overall scale for the free energy of
demixing per unit volume. This captures the coexistence of a fluid
phase in which $\phi = 1$ with an air phase in which $\phi = -1$, with
the two phases separated by a slightly diffuse interface of thickness
$\ell$ and surface tension
\begin{equation}
  \Gamma = \frac{2 \sqrt{2}}{3} G_\mu \ell.
  \label{eq:SI_surface_tension}
\end{equation}
This contributes an additional source term of the form $-\phi \nabla
\mu$ to the Stokesian force balance condition, as discussed in the
main text.  The modulus $G$ and relaxation time $\tau$ that appear in
the viscoelastic constitutive equation are then made functions of
$\phi$, such that viscoelastic stresses only arise in the fluid phase.

Where the fluid meets the hard walls of a flow cell, the boundary
conditions are taken to be~\cite{Yue2010,Dong2012}
\begin{align}
  \vtr{n}\cdot \nabla \mu &= 0,\\
  \vtr{n}\cdot \nabla \phi &= \frac{-1}{\sqrt{2} \ell} \cos{\theta} \left(1 - \phi^2\right).
  \label{eq:SI_CH_BCs}
\end{align}
with $\vtr{n}$ the outward unit vector normal to the wall. The
parameter $\theta$ defines the equilibrium contact angle the interface
between the air and fluid makes with the wall.

At each numerical timestep we first solve the Stokes balance condition
to update the fluid velocity field $\vtr{v}$ at fixed phase field
$\phi$ and polymer stress $\tsr{\Sigma}$, using a streamfunction
formulation to ensure incompressible flow.  We then in turn update the
phase field and viscoelastic stress, with the velocity field
fixed. The advective terms are implemented using a third order
upwinding scheme~\cite{pozrikidis2011introduction}, and any spatially
local terms (which in fact only arise in the viscoelastic constitutive
equation) using an explicit Euler scheme~\cite{numericalrecipes}. To
implement the spatially diffusive terms, in the Lees Edward biperiodic
geometry we use a Fourier spectral method. With walls present, we
instead use a hybrid method: again with Fourier modes in the periodic
vorticity direction $z$, and with finite
differencing~\cite{numericalrecipes} in the flow gradient direction
$y$. All numerical results presented are converged on decreasing mesh
size and increasing mode number.

\end{document}